\documentclass[12pt]{article}
\usepackage{amssymb,amsmath} 
\usepackage[pdftex]{graphicx}
\newcommand{\Letter}{
  \setlength{\textwidth}{7 in}
   \setlength{\textheight}{9.0in}
  \hoffset=-.75in
   \voffset=-1.15in }

\DeclareMathOperator{\Tr}{Tr} 
\DeclareMathOperator{\A}{A}
\DeclareMathOperator{\B}{B}    
\DeclareMathOperator{\holomorphic}{holomorphic} 
\Letter

\begin{document}

\begin{titlepage}
\begin{center}

{\hbox to\hsize{
\hfill SU-ITP-08/06,~SLAC-PUB-13156}}

\vspace{3cm}

{\large \bf Stringy Instantons in IIB Brane Systems\\[2mm]}

\vspace*{1.0cm}

{Shamit Kachru and Du\v san Simi\' c\\[8mm]

\it Department of Physics and SLAC, Stanford University, Stanford, CA
94305/94309}

\vspace*{1.5cm}

{\bf Abstract}\\

\end{center}
\noindent

In this note, we study D-brane instantons which intersect ${\cal N}=1$ supersymmetric
configurations of space-filling D-branes in general Calabi-Yau
compactifications of type IIB string theory.  Our focus is on rigid ``stringy instantons'' -- those which
cannot be interpreted as Yang-Mills instantons in a non-Abelian gauge group on any stack of
space-filling D-branes.  We show that their contributions to the space-time superpotential can
be determined by a topological B-model computation of the tree-level superpotential for
a related auxiliary brane system.  This computation is very tractable as it is governed by an
$A_{\infty}$ structure. We summarize the stringy instanton contribution to the space-time superpotential in a simple formula.



\end{titlepage}
\newpage

\section{Introduction}

\indent \indent D-branes in Calabi-Yau (orientifold) compactifications of type II string theory provide a natural
class of ${\cal N}=1$ supersymmetric quiver gauge theories.  In finding the vacuum structure of these theories (including the possibility of dynamical supersymmetry breaking), or determining the values of various
low-energy couplings of possible phenomenological relevance, it is important to have a handle on non-perturbative contributions to the space-time superpotential.

In addition to conventional field theoretic corrections due to Yang-Mills instantons or strong coupling dynamics, another class of corrections is known to occur in these string models.  Euclidean D-branes
which wrap cycles in the Calabi-Yau space that intersect the space-filling D-branes can contribute
corrections to the superpotential which depend on the charged low-energy fields in an interesting
way.  These corrections have been a focus of recent research, starting with the papers 
\cite{Cvetic,Lust, Uranga,Florea,Herman}.

If the Euclidean branes lie entirely within stacks of multiple space-filling branes, these are just
conventional corrections due to Yang-Mills instantons.   Of more interest are ``stringy instantons''
which wrap distinct cycles in the Calabi-Yau that however intersect the space-filling branes, or
which wrap the same cycle as a single space-filling brane (since these look like $U(1)$ instantons in 4d
and are not conventional low-energy field theory effects).  

So far, the precise rules for computing the spectra and interactions of the open strings connecting
the instantons to the space-filling branes (which are crucial for determining the superpotential correction) have
only been enunciated for toroidal models, orbifolds of tori \cite{italians}, and exactly soluble Gepner points
in Calabi-Yau moduli space \cite{Schellekens}.

For certain slightly more complicated geometries, the contributions have been conjectured on the basis of symmetries and zero mode counting \cite{Florea,franco,seba,shamitofer,eva}. However, a more general treatment is still lacking. In this note we aim to fill this gap by developing the technology for computing the single stringy instanton corrections (up to non-vanishing ${\cal O} (1)$ coefficients) to superpotentials of ${\cal N}=1$ SUSY Type IIB brane configurations on general Calabi-Yau three-folds. 
  
For rigid instantons\footnote{We define  `rigid' stringy instantons to be stringy instantons which have no adjoint moduli other than the four bosonic and two fermionic super-translational modes.} we succeed in reducing the problem to a computation of the tree-level superpotential of a closely related D-brane system, whose 3+1 dimensional gauge theory we refer to as the `auxiliary gauge theory'. We reserve the symbol {\bf{W}} for its superpotential. Such computations lie in the realm of the topological string, and there are powerful methods which make the computation practical for a large class of configurations (which includes all toric and certain non-toric geometries), and possible in principle for any Calabi-Yau three-fold using for example the methods of Aspinwall and Katz \cite{paul} (concretely applied in \cite{lukasz} and extended to orientifolds in \cite{romanians}). The superpotential {\bf{W}} is highly computable due to the ${A}_{\infty}$ structure which governs the
  open topological B-model, as described in those papers.  More concretely, ${\bf{W}}$ is explicitly known for wide classes of brane configurations at toric singularities where gauge dynamics on space-filling D-branes has been studied.  In any of these cases, it follows directly from our results that one can use the known form of ${\bf{W}}$ to compute
stringy instanton corrections to superpotentials for (slightly different) D-brane theories at the same singularities.
We also provide a conjecture for the contribution of non-rigid stringy instantons. 

Our approach is based on exploiting consequences of the ${\cal N} = 2$ super-conformal algebra of the underlying world-sheet description, as well as basic facts about non-perturbative string phenomena at weak coupling.  Therefore, our arguments should hold whenever there is an ${\cal N} = 2$ SCFT description of the D-brane configuration under consideration. 

The organization of this note is as follows. In \S2 we describe the general framework and give a very simple example of our framework in action, while in \S3 we describe our results for rigid stringy instantons.  We turn to non-rigid instantons in \S4, and conclude
in \S5.  For the reader's convenience we present a short appendix summarizing various facts about the 2d ${\cal N} = 2$ super-conformal algebra.

While this work was in progress, we became aware of the related but complementary work appearing in \cite{Riccardo}.

\section{General Framework}

\indent \indent Our space-time is $M_4 \times X$ with X a Calabi-Yau three-fold. We consider ${\cal N}=1$
supersymmetric  collections of space-filling D-branes wrapping even dimensional sub-manifolds of X in type
IIB string theory. There is a 3+1 dimensional field theory living on the world-volume of the D-brane system and we are interested in computing the corrections to its superpotential due to stringy instantons.

Imagine organizing the D-brane configuration in terms of a quiver. The world-volume gauge theory will in general receive contributions from Euclidean branes wrapping each of the quiver's different nodes. Throughout this paper we focus on instantons wrapping unoccupied nodes (more correctly, nodes with trivial gauge dynamics -- we treat the case of an instanton in a $U(1)$ gauge group with only massive matter as well). We refer to these instantons as `stringy' to distinguish them from those wrapping occupied nodes which in general have gauge theory counterparts.\footnote{In some brane systems which admit multiple UV completions, e.g. by both a renormalization group cascade and a direct coupling to 10d string theory, the same effects may be derivable in two different ways as gauge theory effects and stringy effects respectively \cite{shamitofer}.}

Let $S_{eff}$ be the effective action of the instanton; it encodes the interactions of the instanton moduli with themselves and the world-volume fields of the quiver gauge theory. Then the contribution to the gauge theory action from a single instanton is
  
\begin{eqnarray}
\label{master}
\Delta S =  \int d\mu_i   e  ^{-S [\mu_i, \Phi_i, \bar \Phi_i]_{eff}} + c.c ,
\end{eqnarray}
where the $c.c.$ arises from the anti-instanton,
$\mu_i$ collectively denotes all of the instanton moduli, and $\Phi_i$ and $\bar \Phi_i$ collectively denote the gauge theory superfields which couple non-trivially to the $\mu_i$. This prescription, in addition to being consistent with semi-classical reasoning, coincides with the correct prescription for gauge theory instantons (see for example \cite{gauge,italians}). From the stringy point of view, gauge theory instantons and stringy instantons are on equal footing, and so we assert that (\ref{master}) is the correct prescription even for the stringy case. 
 
We are interested in corrections to the superpotential so we restrict to the case where the instanton has only two super-translational fermi modes.  This condition can be easily met for instantons which either intersect O-planes
\cite{bianchi,Schellekens,italians,seba}, or instantons
which lie in $U(1)$ gauge factors of the space-time theory (for a general reference on O-planes,
see \cite{hori}). Then (\ref{master}) reduces to


\begin{eqnarray}
\label{general}
\Delta S = \int d^4x d^2\theta \left ( \int d\mu_i'   e  ^{-S [\mu_i', \Phi_i, \bar \Phi_i]_{eff}} \right) + c.c.,
\end{eqnarray} 
where we've explicitly separated the super-translational modes from the other moduli, denoted by $\mu_i'$.  The correction to the space-time superpotential, is obtained by taking the holomorphic part of (\ref{general}), 

\begin{eqnarray}
\label{deltaW}
\Delta W =  \Biggl (  \int d\mu_i'   e  ^{-S [\mu_i', \Phi_i, \bar \Phi_i]_{eff}} \Biggr )_{\holomorphic}~.
\end{eqnarray} 
One might expect (\ref{general}) to automatically pull out only the conventional superpotential; we have made the specification (\ref{deltaW}) only to distinguish between regular superpotential terms and multi-fermion F-terms
\cite{beasley}, which can be computed in the same formalism \cite{weigand,park} and are not manifestly
holomorphic.

\begin{figure} [top]
 \centering 
 \includegraphics[width=0.6\textwidth]{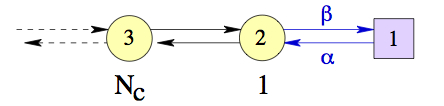} 
 \caption{A typical extended quiver. The Euclidean brane wraps the square node.} 
 \label{quiver} 
\end{figure}

As an illustration, consider the orientifolded-orbifolded-conifold model of \cite{franco,shamitofer}. The quiver is depicted in Fig. \ref{quiver}. Those authors conjectured that the instanton couples to the world-volume theory through a pair of fermions, $\alpha$ and $\beta$ (referred to as `Ganor' strings \cite{ori}), and that the instanton effective action takes the form

\begin{eqnarray}
\label{special}
S_{eff} = t +  c ~ \alpha \Phi_{32} \Phi_{23} \beta + \dots ,
\end{eqnarray} 
where $c$ is some constant and $t$ is the (complexified) volume of the cycle wrapped by the D-instanton.
This leads to a correction to the space-time superpotential:

\begin{eqnarray}
\label{shamitofer}
\Delta W \sim e^{-t} \left( \Phi_{32} \Phi_{23} + \cdots \right)~.
\end{eqnarray} 

The derivation of this expression in our framework is simple and would proceed as follows. Replace the instanton with a single space-filling D-brane; the 3+1 dimensional gauge theory we obtain in this way is referred to as the `auxiliary gauge theory'. Compute \textbf{W}, the tree-level superpotential for this gauge theory. The answer is \cite{shamitofer}

\begin{eqnarray}
\textbf{W} =\Tr A \Phi_{32}\Phi_{23} B + f(\Phi_{ij})~,
\end{eqnarray} 
where $A, B$ are chiral multiplets stretching between nodes 1 and 2 (replacing $\alpha$ and $\beta$ in the extended quiver above), and $f(\Phi_{ij})$ is independent of A and B. 
Using {\bf{W}}, extract the coefficients $C_{ijl_1\cdots l_n}$


\begin{eqnarray}
C_{i j l_1\cdots l_n } = d(l) {\partial^{n+2} \textbf{W} \over \partial \A_i \partial \B_j  \partial \Phi_{l_1} \cdots \partial \Phi_{l_n}} \Big \vert_{\A, \B, \Phi = 0}
\end{eqnarray}
where $d(l)$ is a combinatorics factor; for example $d(l) = {1 \over n!}$ when $l_1 = \cdots = l_n$. The correction to the 4d superpotential is then given by

\begin{eqnarray}
\label{deltaW3}
&\Delta W \propto  e^{-t} \det M(\Phi) \nonumber \\ \\
&M(\Phi)_{ij} \equiv  \left( \displaystyle \sum^{\infty}_{n=1} \displaystyle C_{i j l_1\cdots l_n }  \Phi_{l_1} \cdots \Phi_{l_n} \right ) \nonumber
\end{eqnarray}
\begin{eqnarray}
& \Rightarrow \Delta W \propto e^{-t} \Phi_{32} \Phi_{23} ~.
\end{eqnarray}
We note here that implicit in ${\bf{W}}$ is an overall prefactor which is an unknown, but constrained function of
the K\"ahler moduli (whose leading term is just a non-vanishing constant), as discussed in \cite{Brunner}.  This prefactor of course carries over into the computation of $\Delta W$, but will remain implicit in the rest of our
discussion.



\section{The Rigid Case}

\indent \indent In this section we discuss the case of rigid instantons.  In \S3.1 we analyze the instanton's moduli and prove that the charged moduli consist entirely of Ramond ground-states of the world-sheet ${\cal N}=2$ SCFT; consequently there are no charged bosonic zero modes. In \S3.2, we describe the GSO projection. In \S3.3, we compute the instanton's effective action, and we derive a formula for the instanton's contribution to the 4d superpotential in \S3.4. Finally, in \S3.5, we extend the results to instantons which lie in massive $U(1)$ factors (i.e. single space-filling D-branes with no massless adjoint matter).

\subsection{Moduli Spectrum}

\indent \indent The rigidity means that the adjoint moduli consist exclusively of the four bosonic and two fermionic super-translational modes, and so here we focus on the sector of strings (called the `charged' sector) stretching between the instanton and a space-filling brane. The mass of a particular state in the Ramond sector is

\begin{eqnarray}
\label{massR}
m^2_{i} = -{5 \over 8} + 4 \times {1 \over 16} + \Delta_{i},
\end{eqnarray}
where the first contribution comes from the ghosts, the second from the $3+1$ ``space-time" directions, and the third represents the weight of the corresponding state in the internal ${\cal N} = 2$ SCFT.  

Setting $m^2_{i} = 0$ implies $\Delta_{i} = {3 \over 8}$. Ramond states of this weight are special in the underlying ${\cal N}=2$ SCFT. They correspond to ground-states of the Ramond sector, and are related to chiral/anti-chiral primaries in a one-to-one manner via spectral flow. These are the `Ganor strings' assumed by previous authors.  Ganor strings are topological and their spectrum is computable in a wide variety of cases  \cite{paul,romanians,lerchevafa,greene}.

In the NS sector, we have
\begin{eqnarray}
\label{massNS}
m^2_{i} = -{1 \over 2} + 4 \times {1 \over 8} + \Delta_{i},
\end{eqnarray}
and so charged bosonic moduli must arise from states with $\Delta_{i} = 0$. However, such a state would contradict the unitarity of the underlying CFT (under the assumption that the stringy instanton is wrapping a node of the quiver which isn't wrapped by any space-filling D-brane; we postpone the discussion of nodes occupied by `massive' U(1) gauge factors until \S3.5).

\begin{figure}[top]
\centering 
\includegraphics[width=0.6\textwidth]{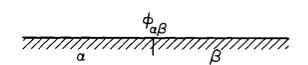} 
\caption{A discontinuity in the boundary conditions corresponds to the insertion of a local operator.} 
\label{bc} 
\end{figure}

To see this, consider the internal CFT on the upper-half plane. The negative real axis has boundary conditions $A$ (corresponding to the end of the Ganor string on the instanton), whereas the positive real axis has boundary conditions $B$ (the end of the string on the space-filling brane).  Since the change in boundary conditions occurs at a point (the origin) it can be represented by a local operator, $\phi_{AB}$, as in Fig. \ref{bc} $\cite{cardy}$. The NS ground-state energy of the CFT is thus given by the weight of the operator which creates the boundary conditions $AB$ from the ``true" vacuum (ie: the groundstate of the CFT with just AA or BB boundary conditions). In a unitary CFT, states which are not the identity operator  necessarily have positive-definite weight. Since we assume the instanton and D-brane are wrapping different cycles, it must be that $\phi_{AB}$ is not the identity operator. Hence $\Delta_{i} > 0$ for all states in the NS sector and we see that there 
 are no charged bosonic moduli.

\subsection{GSO Projection}

\indent \indent The Ganor strings can be divided evenly into two groups; those related to chiral primaries and those related to anti-chiral primaries (under spectral flow). It turns out that the GSO projection will preserve one group while annihilating the other. The two possibilities correspond to instanton versus anti-instanton. To see that this is the correct GSO projection, consider the theory in which the instanton is replaced by a space-filling brane of the same type. The vertex operators for the (analogues of the) Ganor strings then have the form

\begin{eqnarray}
V_{\alpha I}(z) =  S_{ \alpha} \Sigma_I(z)  e^{ik \cdot X (z)}e^{-{\phi(z) \over 2}}, \quad \bar V^{\dot \alpha}_I(z) =  S^{ \dot \alpha} \bar \Sigma_I(z) e^{ik \cdot X (z)} e^{-{\phi(z) \over 2}}
\end{eqnarray}
where $S^{\dot \alpha}$ and $S_{\alpha}$ are the usual spin fields creating Weyl spinors of opposite chirality in the 3+1 Minkowski dimensions, while $\Sigma_I$ and $\bar \Sigma_I$ are the spin fields associated to the chiral and anti-chiral Ganor strings, respectively. In string theory, $S^{\dot \alpha}$ and $S_{\alpha}$ have opposite charge under the GSO projection, from which it follows that the $\Sigma_I$ and $\bar \Sigma_I$ are oppositely charged under the GSO projection as well. Thus, for the instanton we keep the Ganor strings, while projecting out the anti-Ganor strings, and vice-versa for the anti-instanton. 

\subsection{Instanton Effective Action}

\indent \indent To compute the effective action we take all the massless modes present in the system and compute all the possible tree-level scattering amplitudes between them. Let $\alpha_{i}$ and $\beta_{i}$ denote the Ganor strings (of opposite orientations, as in Fig. \ref{quiver}), and let $\Phi_{i}$ denote the chiral superfields of the world-volume theory. As we discussed earlier, for an instanton the GSO projection picks out the Ganor strings related via positive spectral flow to chiral primaries.  It follows that the Ganor strings will carry R-charge $-{1 \over 2}$ under the U(1) associated with the ${\cal N} =2$ super-conformal algebra. The (vertex operators for the) $\Phi_i$, on the other hand, carry R-charge $+1$.  A sketch of the R-charge counting can be found in appendix A. 

Only amplitudes with a net R-charge of zero can yield a non-zero contribution. On the disk, the only amplitudes that respect this condition and are holomorphic in the $\Phi_i$ are of the form

\begin{eqnarray}
\label{quad}
A(z) \langle  \alpha_{i}(z_i) \beta_{j} (z_j) \Phi_{l_1} (z_{l_1}) \int _{ \partial D} G_{-{1 \over 2}} \cdot \Phi_{l_2} \cdots \int _{ \partial D} G_{-{1 \over 2}} \cdot \Phi_{l_{n}}\rangle
\end{eqnarray}

\begin{eqnarray}
\label{random1}
A(z) = \langle \Delta(z_i) \bar \Delta (z_j) c(z_i)e^{-{\phi \over 2}(z_i)} c(z_j)e^{-{\phi \over 2}(z_j)} c(z_{l_1})e^{-\phi(z_{l_1})} \rangle,
\end{eqnarray}
where $\alpha_{i}$ and $\beta_{i}$ are the spin-fields representing the Ganor strings in the internal CFT,  $\Delta$ and $\bar \Delta$ are the twist fields which create the mixed Neumann-Dirichlet boundary conditions in the 3+1 Minkowski directions, and $e^{-l \phi}$ and $c$ are the usual ghost vertex operators and ghost insertions, respectively.
Here we are working at fixed values of the closed string moduli; more generally,
the B-model could also in principle
determine the complex-structure dependence of the open string amplitudes.



The amplitudes in (\ref{quad}) can be extracted from the superpotential of a closely related system. Consider the D-brane system in which the instanton is replaced by a space-filling brane. We refer to this system as the `auxiliary system' and denote the superpotential of the 4d gauge theory that it produces by \textbf{W}. The quantity analogous to (\ref{quad}) is a fermion bilinear scattering amplitude:

\begin{eqnarray}
\label{random2}
&B(z) \langle  \alpha_{i}(z_i) \beta_{j} (z_j) \Phi_{l_1} (z_{l_1}) \int _{ \partial D} G_{-{1 \over 2}} \cdot \Phi_{l_2} \cdots \int _{ \partial D} G_{-{1 \over 2}} \cdot \Phi_{l_{n}}\rangle =  d(l) {\partial^{n+2} \textbf{W} \over \partial \A_i \partial \B_j  \partial \Phi_{l_1} \cdots \partial \Phi_{l_n}} \Big \vert_{\A, \B, \Phi = 0}
\end{eqnarray}

\begin{eqnarray}
&B(z) \propto \epsilon_{\dot \alpha \dot \beta} \langle S^{\dot \alpha}(z_i)  S^{\dot \beta} (z_j) c(z_i)e^{-{\phi \over 2}(z_i)} c(z_j)e^{-{\phi \over 2}(z_j)} c(z_{l_1})e^{-\phi(z_{l_1})} \rangle.
\end{eqnarray}
where $d(l)$ is a combinatorics factor. For example $d(l) = {1 \over n!}$ when $l_1 = \cdots = l_n$. The relation $A(z) \propto B(z)$ is easily seen \cite{gauge} and the constant of proportionality is the same for all possible choices of $\alpha$'s, $\beta$'s and $\Phi$'s in (\ref{random1}) and (\ref{random2}), so we may ignore it. Hence, the effective action is

\begin{eqnarray}
\label{theaction}
S_{eff} = t +   \left( \sum^{\infty}_{n=1} C_{i j l_1\cdots l_n } \alpha_i \Phi_{l_1} \cdots \Phi_{l_n} \beta_j \right ) + \dots  
\end{eqnarray}

\begin{eqnarray}
\label{C}
C_{i j l_1\cdots l_n } = d(l) {\partial^{n+2} \textbf{W} \over \partial \A_i \partial \B_j  \partial \Phi_{l_1} \cdots \partial \Phi_{l_n}} \Big \vert_{\A, \B, \Phi = 0}~.
\end{eqnarray}
The ellipses denote terms non-holomorphic in $\Phi_i$ which are irrelevant for determining superpotential
corrections (but may be relevant for computing, e.g., multi-fermion F-terms \cite{beasley}).  
It would be interesting to explore these further.

\subsection{The Superpotential Correction}

\indent \indent Computing the correction to the 4d superpotential is now the work of a moment. Plugging (\ref{theaction}) into (\ref{deltaW}) we arrive at:

\begin{eqnarray}
\label{deltaW2}
\Delta W \propto  e^{-t} \det M(\Phi)
\end{eqnarray}

\begin{eqnarray}
\label{M}
M(\Phi)_{ij} \equiv  \left( \displaystyle \sum^{\infty}_{n=1} \displaystyle C_{i j l_1\cdots l_n }  \Phi_{l_1} \cdots \Phi_{l_n} \right ) 
\end{eqnarray}
where the $C_{ij l_1 \cdots l_n}$ were given in (\ref{C}) and the constant of proportionality in (\ref{deltaW2}) is a determinant of massive modes which arises from expanding the DBI action around the instanton background.  Being a determinant of massive modes (which follows from the rigidity of the instanton), it is non-vanishing. There is also a factor of the string scale to some appropriate power, determined by dimensional analysis. 

\subsection{Wrapping a Massive U(1) Node}

\indent \indent Thus far we have we have left unspecified the mechanism which removes the $\bar \theta$'s. The typical procedure is to introduce an orientifold plane cutting through the instanton in such a way that it does not introduce adjoint matter (this latter requirement is obviously removed if one is willing to consider non-rigid D-instantons). This may not always be possible, or phenomenologically viable, so it is nice to have an alternative. In \cite{petersson} it was shown that for the case of orbifolds one could lift the $\bar \theta$'s by having the instanton wrap a node occupied by a massive U(1).\footnote{`Massive' refers to the fact that the cycle is rigid. Since U(1) factors do not give rise to gauge theory instantons these instantons still fall into the `stringy' category.}
In \cite{aganagic}, geometric transitions on such massive $U(1)$ nodes were also used to compute stringy instanton effects (and sum infinite series of instantons), in slightly more elaborate geometries.
We now show that the direct instanton analysis of \cite{petersson} for massive $U(1)$ nodes extends to the more general D-brane configurations studied in this paper. 

Let us reconsider the spectrum of instanton moduli discussed in \S3.1. There is now a new sector of strings, those stretching between the stringy instanton and the massive U(1). In the NS sector, we see that the unitarity argument no longer forbids the existence of massless bosonic states. This is because the U(1) and the instanton wrap the same node and so $\phi_{AB} \propto  1_{AB}$. By unitarity of the internal CFT, a state with $\Delta_i = 0$ is necessarily the NS sector ground-state.  There are two such states, one for each orientation of a string stretching between the U(1) and the stringy instanton. Because of the fact that the 3+1 space-time directions have mixed (Neumann-Dirichlet) boundary conditions, the space-filling fermionic oscillators have integral mode expansions in the NS sector. Hence, these charged bosonic moduli carry spinor indices. We denote the charged bosons surviving the GSO projection by $\omega^{\dot \alpha}$ and $\tilde \omega^{\dot \alpha}$ (tilde denotes a reversal of orientation). 

In the Ramond sector the only ground-states are those related by spectral flow to the Neveu-Schwarz ground-state (since we are assuming the node is rigid). For each orientation of strings there are two of these, one arising from positive spectral flow the other via negative spectral flow. However, the GSO projection eliminates one such state for each orientation. We call the resulting states $\mu$ and $\tilde \mu$. Since the vertex operators for the $\omega$'s and $\mu$'s don't depend on the details of the internal CFT, the instanton's action has the following universal structure:

\begin{eqnarray}
&S = i \left (\tilde \mu \omega_{\dot \alpha}  + \tilde \omega_{\dot \alpha} \mu \right ) \bar \theta^{\dot \alpha} -iD^c \left ( \tilde \omega^{\dot \alpha} (\tau^c)^{\dot \beta}_{\dot \alpha} \omega_{\dot \beta} \right ) + {1 \over 2} \tilde \omega _{\dot \alpha} f(\Phi, \bar \Phi) \omega^{\dot \alpha} + \nonumber \\ \\ &   t +   \left( \sum^{\infty}_{n=1} C_{i j l_1\cdots l_n } \alpha_i \Phi_{l_1} \cdots \Phi_{l_n} \beta_j \right ) + \dots  \nonumber
\end{eqnarray}
where the $\tau^c$ are Pauli matrices and $D^c$ is a Lagrange multiplier field whose purpose is to implement the ADHM type constraint $\ \tilde \omega^{\dot \alpha} (\tau^c)^{\dot \beta}_{\dot \alpha} \omega_{\dot \beta} = 0$  \cite{gauge}. Integrating over the $\omega$, $\tilde \omega$, $\mu$, $\tilde\mu$, $D$, and $\bar \theta$'s contributes an ${\cal{O}}(1)$ constant (it is independent of the $\bar \Phi$'s):

\begin{eqnarray}
&\Delta S =  \int d^4x d\{\theta, \bar \theta, \mu, \bar\mu, \omega, \bar\omega, D^c, \alpha_i, \beta_i \}  e ^ {-S_1 - S_2} + c.c. \nonumber \\  \\ 
&\propto \int d^4 xd^2 \theta \left ( \int d\{ \alpha_i, \beta_i \} e^{-S_{eff}[\alpha_i, \beta_i, \Phi_i, \bar \Phi_i]} \right ) +c.c.  \nonumber 
\end{eqnarray}
with $S_{eff}$ as in (\ref{theaction}). Hence, the sole effect of the massive U(1) is to lift the extra $\bar \theta^{\dot \alpha}$ zero modes, and the resulting contribution to the spacetime superpotential is given again by (\ref{deltaW2}).


\section{The Non-Rigid Case}

\indent \indent We now turn to the case where the stringy instanton has additional light adjoint matter on its world-volume. We call these adjoint matter fields $X_a$ and their Grassmann superpartners $\chi_a^{\dot \alpha}$ (as distinct from the Ganor strings $\alpha, \beta$). There is a natural conjecture generalizing our result for rigid instantons, which however involves a non-trivial correlation function in the instanton world-volume theory.

\subsection{Instanton Effective Action}
\indent \indent For computing the superpotential correction, there are two relevant types of amplitudes that could be non-vanishing, consistent with the worldsheet
R-charge constraints:

\begin{eqnarray}
&\langle \alpha_i (z_i) \beta_j (z_j) \Phi_{l_1} (z_{l_1})  \int_{ \partial D} G_{-{1 \over 2}} \cdot \Phi_{l_2} \cdots \int _{ \partial D} G_{-{1 \over 2}} \cdot \Phi_{l_{n}} \int _{ \partial D} G_{-{1 \over 2}} \cdot X_{r_{1}} \cdots \int _{ \partial D} G_{-{1 \over 2}} \cdot X_{r_{m}} \rangle 
\end{eqnarray}


\begin{eqnarray}
&\langle \alpha_i (z_i) \beta_j (z_j) \bar \chi_a^{\dot \alpha} (z_{a})  \int _{ \partial D} \bar \chi^{\dot \beta}_b  \int _{ \partial D} {\cal{O}}(X, \bar X) \int _{ \partial D} G_{-{1 \over 2}} \cdot \Phi_{r_{1}} \cdots \int _{ \partial D} G_{-{1 \over 2}} \Phi_{r_{n}}  \rangle
\label{offensive}  
\end{eqnarray}
where the $\int _{ \partial D} {\cal{O}}(X, \bar X)$ denote schematically some string of integrated vertex operators composed purely of the $X_a$ and $\bar X_a$'s. 

Both classes of correlators can be connected with related computations in the auxiliary gauge theory.  The only terms which are expected to be holomorphic in all space-time chiral superfields (including the closed string moduli) are those of the first type, which are also computable in the topological B-model (for the auxiliary brane configuration). Then the natural conjecture is that the superpotential correction is still computable in terms of ${\bf{W}}$, as follows:

\begin{eqnarray}
\label{deltaWadjoint}
\Delta W \propto  e^{-t} \det M(\Phi)
\end{eqnarray}

\begin{eqnarray}
M_{ij} =  \displaystyle{\sum_{n+m \geq 1} }C_{ij k_1 \cdots k_n l_1 \cdots l_m}\Phi_{k_1} \cdots \Phi_{k_n} \langle X_{l_1} \cdots X_{l_m}  \rangle 
\end{eqnarray}

\begin{eqnarray}
\label{CX}
C_{i j l_1\cdots l_n } = d(k,l) {\partial^{n+m+2} \textbf{W} \over \partial \A_i \partial \B_j  \partial \Phi_{k_1} \cdots \partial \Phi_{k_n}  \partial X_{l_1} \cdots \partial X_{l_m}  }\Big \vert_{\A, \B, \Phi, X = 0}
\end{eqnarray}
 where $d(k,l)$ is the usual combinatorics factor as in (\ref{C}) and $\langle X_{l_1} \cdots X_{l_m}  \rangle$ is a correlation function in the world-volume theory on the instanton. For e.g. D1-instantons, the world-volume theory is in general (assuming the adjoint deformations are obstructed) some Landau-Ginzburg theory; for unobstructed D1-branes it is instead a non-linear sigma model on the moduli space of the relevant curve.  Note that the path integral evaluating $\langle X_{l_1} \cdots X_{l_m}\rangle$  includes integrals over the modulini $\chi_{a}^{\alpha}$; if they cannot be absorbed by
 pulling down factors from the action, it will vanish.

\section{Discussion}

\indent \indent In this note, we have shown using elementary world-sheet CFT techniques that the superpotential corrections induced by rigid stringy instantons to quiver gauge theories on space-filling D-branes can be related simply to the superpotential of an auxiliary gauge theory arising on a slightly different brane system.  The superpotential of the auxiliary theory is highly computable using the ${A}_{\infty}$ structure of the topological B-model as in \cite{paul}.  This provides a very effective technique for computing stringy instanton corrections.

In particular, for any D-brane system which arises at e.g. a toric Calabi-Yau singularity with a known quiver and
interactions, one can immediately use our results to read off the effects induced by rigid stringy instantons.  Our
result confirms the analyses of \cite{franco,seba,shamitofer,eva} in special cases, but applies much more generally.
Since general Calabi-Yau manifolds provide a much richer territory for model building than simple toroidal orientifolds, one can hope that these results will be of use in concrete applications.

\bigskip \bigskip

\centerline{\bf{Acknowledgements}}

\bigskip
We would like to thank Allan Adams, Bogdan Florea, Hans Jockers, Albion Lawrence, Alessandro Tomasiello and Timo Weigand for helpful discussions.  
We are also grateful to Riccardo Argurio, Gabriele Ferretti and Christoffer Petersson for sharing with us their ideas and
their draft with many related results before publication.  This work was supported by the
Stanford Institute for Theoretical Physics, the NSF under grant PHY-0244728, and the
DOE under contract DE-AC03-76SF00515.

\appendix 

\section{Facts about ${\cal N}=2$ SCFT }

\indent \indent 
Here we summarize some useful facts about operators in 2d ${\cal N} = 2$ SCFT; useful references for this material are \cite{lerchevafa, greene}. Consider a state $\Psi_{h,q}$, where $h$ and $q$ denote it's weight and R-charge respectively. Let us act on this state with $\eta$ units of the spectral flow operator. Then the resulting state is some $\tilde \Psi_{h_\eta,q_\eta}$ with

\begin{eqnarray}
h_\eta = h - \eta q + {c \over 6} \eta^2, \quad q_\eta = q - {c \over 3} \eta~,
\end{eqnarray}
where $c$ is the central charge of the conformal field theory. The case with $\eta = \pm {1 \over 2}$ corresponds to spectral flow which maps the Neveu-Schwarz sector into the Ramond sector, and is the only one of interest here. For $c = 9$ (which is the case for a sigma-model whose target is a Calabi-Yau  three-fold) we have

\begin{eqnarray}
h_{\pm 1/2} = h \mp {q \over 2} + {3 \over 8}, \quad q_{\pm 1/2} = q \mp {3 \over 2}~.
\end{eqnarray}

Thus if $\Psi_{h,q}$ is a chiral (anti-chiral) primary then under spectral flow with  $\eta =  +{1 \over 2}$ ($\eta =  - {1\over2}$) it gets mapped to a Ramond ground-state with weight $+ {3 \over 8}$ and R-charge $q- {3\over2}$ ($q +{3\over 2}$). In our case, the relevant chiral primaries all have $h = + {1\over2}$, as these are the ones that correspond to massless states in the auxiliary gauge theory. Thus, their R-charge is +1 and the R-charges of the corresponding Ramond ground-states are $-{1\over2}$. This explains some of the R-charge counting of \S3. In addition the R-charge of the integrated vertex operators $G_{-{1\over2}} \cdot \Phi$ is zero, since

\begin{eqnarray}
G_{-{1\over2}} \cdot \Phi = (G^+_{-{1\over2}} + G^-_{-{1\over2}}) \cdot \Phi = G^{-}_{-{1 \over 2}} \cdot \Phi~.
\end{eqnarray}
Here we've used the fact that $G^+_{-{1\over2}} \cdot \Phi = 0$ (since $\Phi$ is a chiral primary). Using the fact that $G^{-}$ has R-charge $-1$ we see that  $G_{-{1\over2}} \cdot \Phi$ has R-charge zero.

\newpage


\end{document}